\documentclass[a4paper,11pt]{article}
\usepackage{pos}

\title{Design of an Imaging Air Cherenkov Telescope array layout with differential programming}
\ShortTitle{Design of an IACT array layout with differential programming}

\author*[a]{Cyril Alispach}
\author[a]{Matthieu Heller}
\author[a]{Teresa Montaruli}

\onbehalf{for the SST-1M Collaboration}

\affiliation[a]{D\'epartement de Physique Nucl\'eaire, Facult\'e de Sciences, Universit\'e de Gen\`eve, 24 Quai Ernest Ansermet, CH-1205 Gen\`eve, Switzerland}

\emailAdd{cyril.alispach@unige.ch}

\abstract{
Current optimization of ground-based Cherenkov telescopes arrays, also called Imaging Air Cherenkov Telescope (IACT) arrays, relies on brute-force human-driven approaches based on large simulations requiring both high amount of storage and long computation time. To explore the full phase space of telescope positioning of a given array even more simulations would be required. To optimize any array layout, we explore the possibility of developing a differential program with surrogate models of IACT arrays based on high-level instrument response functions (IRFs).

The simulation time of a single telescope to a cosmic-ray event can be significantly reduced with its instrument response function or with generative models. However, it is not straight forward to model the array of telescopes from a set of single telescope surrogate models as the array is a stereoscopic imaging system. The complexity increases as well if the telescopes in the array are of different types.

Additionally, the optimum of the array layout depends on the scientific use case. Current array layout optimization are obtained by minimizing the sensitivity of the array, a metric that depends on several high-level parameters such as the trigger efficiency, the energy and angular resolution, as well as the background rejection capability. The variety of telescopes types in IACT arrays, such as in the Cherenkov Telescope Array Observatory (CTAO), not only extends the sensitive energy range but also allows for cross-calibration of the instruments. Therefore, the optimal array layout is not only which minimizes sensitivity but also which reduces the systematic uncertainties.

We focus on the optimization of a telescope array based on the SST-1M IACTs in Hanle, Ladakh India aiming at building a generic optimization pipeline for future ground-based cosmic-ray observatories.}

\FullConference{Fifth MODE Workshop on Differentiable Programming for Experiment Design (MODE2025)\\
8-13 June 2025\\
Kolymbari, Crete, Greece\\}

\begin{document}
\maketitle

\section{Introduction: array layout optimization of ground-based gamma-ray astronomy with IACTs}

High-energy gamma rays entering the Earth's atmosphere initiate electromagnetic cascades, producing secondary particles—electrons, positrons, and photons. Charged particles traveling faster than the speed of light interact with the medium that emits Cherenkov radiation, faint flashes of blue-UV light. This light is emitted in a narrow cone (typically 0.1°–1.4°), forming a “light pool” on the ground with a radius of 120–150 meters and relatively uniform photon density.
These flashes last only a few nanoseconds and produce just a few photons per square meter. Cherenkov telescopes use fast, UV-sensitive cameras with high photon detection efficiency to capture them. With effective areas ranging from $10^5$ to $10^6$\,m², depending on energy, IACTs can detect gamma rays from tens of GeV up to 100\,TeV, making them powerful tools for studying extreme astrophysical phenomena.


Traditional optimization of IACT array layouts relies on brute-force Monte Carlo simulations using \textit{CORSIKA}~\cite{CORSIKA} and \textit{sim\_telarray}~\cite{simtel}, which are computationally expensive and require large storage. Exploring the full configuration space of telescope positions would demand even more resources.
To address this, we propose a differentiable modeling approach using surrogate models based on high-level instrument response functions. While single-telescope simulations can be accelerated using such models, extending them to arrays is non-trivial due to the stereoscopic nature of the system.

Previous layout studies, such as in~\cite{CTAlayout}, rely on extensive simulations but lack efficiency in exploring the full parameter space. In contrast, differentiable techniques—like those used in~\cite{SWIGO} for the SWGO water Cherenkov array—enable scalable, gradient-based optimization.
In this work, we present a differentiable model for estimating the stereoscopic performance of arbitrary array layouts, derived from monoscopic telescope performance.

\begin{figure}
    \centering
    \includegraphics[width=0.63\linewidth]{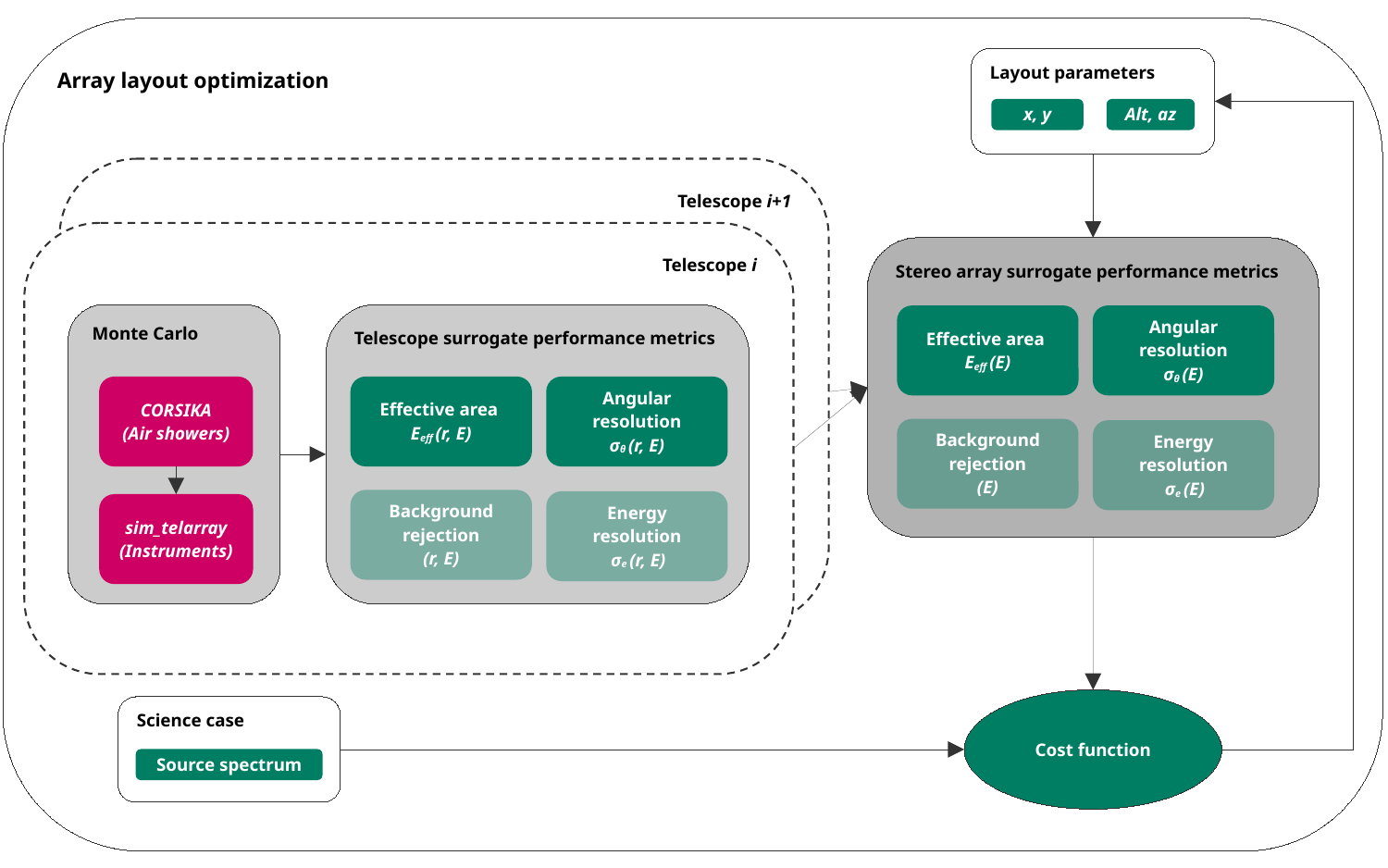}
    \caption{Schematic of the array layout optimization workflow.}
    \label{fig:scheme}
\end{figure}

Although the method applies to arrays of any size and telescope type, we focus on a simplified case: two identical SST-1M telescopes~\cite{SST-1M} separated by a distance $\Delta x$. In Section~\ref{sec:optimization}, we define a cost function to find the optimal separation for detecting a gamma-ray source with spectral index $\Gamma = -2$. The optimization scheme is illustrated in Figure~\ref{fig:scheme}. 

\section{Imaging Atmospheric Cherenkov Telescope Performance Metrics}

IACT performance is typically characterized by four key metrics:
\begin{itemize}
    \item \textbf{Effective area} – the ground area over which the instrument is sensitive to gamma-ray events.
    \item \textbf{Angular resolution} – the precision of direction reconstruction.
    \item \textbf{Energy resolution} – the accuracy of energy reconstruction.
    \item \textbf{Gamma/hadron separation} – the ability to distinguish gamma rays from hadronic background.
\end{itemize}

An additional metric, the \textbf{differential flux sensitivity}, combines the above and quantifies the instrument’s ability to detect a given flux spectrum under predetermined exposure and significance thresholds. It is considered the gold standard for layout optimization, as done in~\cite{CTAlayout}.

In this study, we focus on effective area and angular resolution. While not sufficient alone to define an optimal layout, they provide a strong basis for maximizing detection rates and image sharpness.
The stereoscopic effective area and angular resolution are derived from monoscopic metrics in Sections~\ref{sec:aeff} and~\ref{sec:angular_resolution}, respectively. All metrics are evaluated as functions of gamma-ray energy from 10 GeV to 100 TeV, using five bins per energy decade, following community standards.
Energy resolution and gamma/hadron separation are not addressed here but will be considered in future work (see Section~\ref{sec:conclusion}).

Gamma-ray simulations for the SST-1M telescope were performed using \textit{CORSIKA} + \textit{sim\_telarray}, with $10^9$ on-axis primary gamma rays simulated at a zenith angle of $20^\circ$. The energy range spans 10\,GeV to 100\,TeV, following a power-law spectrum with index $E^{-2}$. Simulations were conducted for the Hanle Observatory (altitude 4500\,m), using the corresponding atmospheric profile and geomagnetic field. For the moment, we optimized only on signal and did not yet use the background.
 
\subsection{Effective Area}\label{sec:aeff}

The effective area of an IACT, both monoscopic and stereoscopic, can be derived from the survival (events that survive the trigger and the analysis cuts) probability of the telescopes.

\subsubsection{Monoscopic Effective Area}

\begin{figure}
    \centering
    \includegraphics[width=0.3\linewidth]{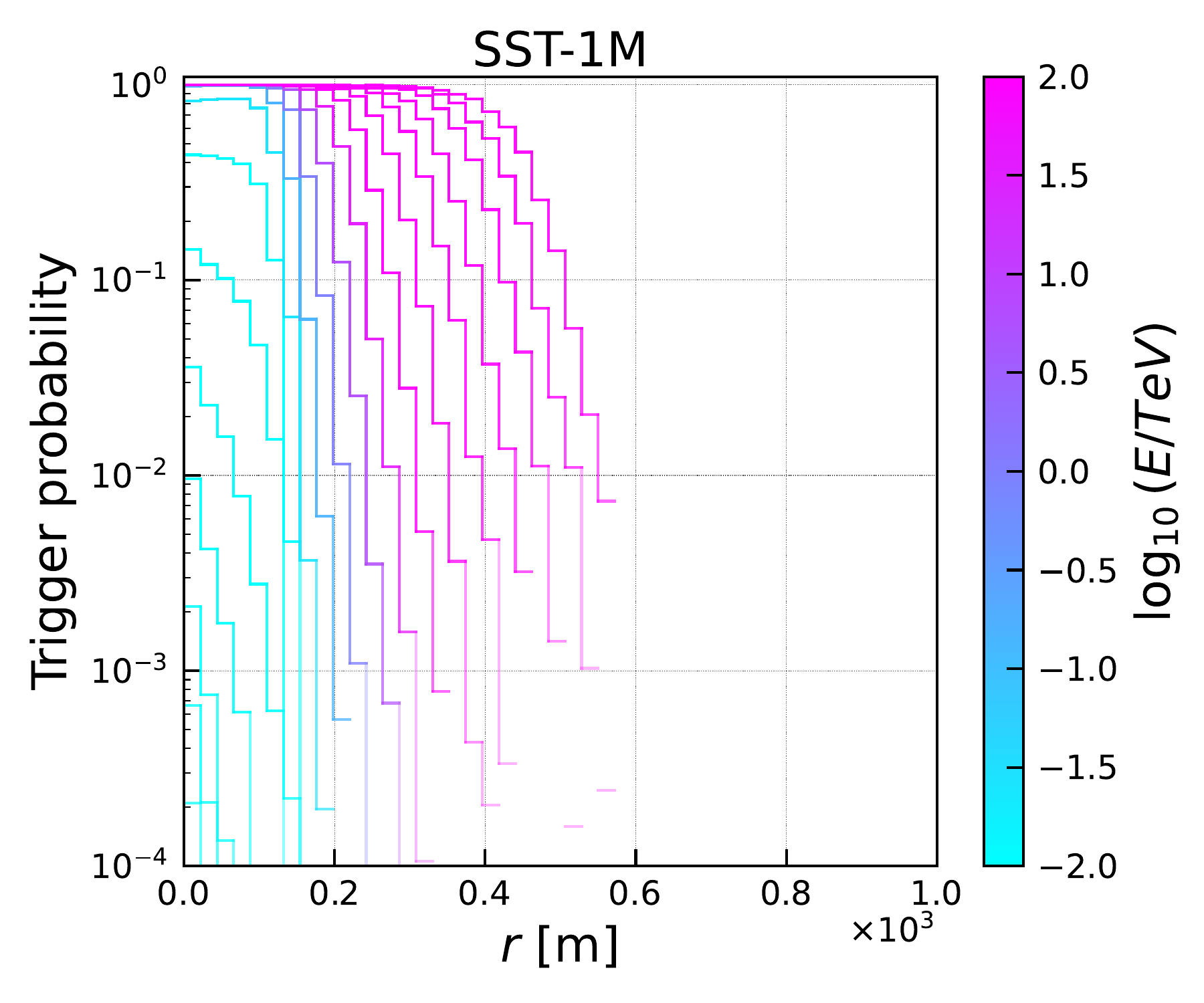}
    \includegraphics[width=0.3\linewidth]{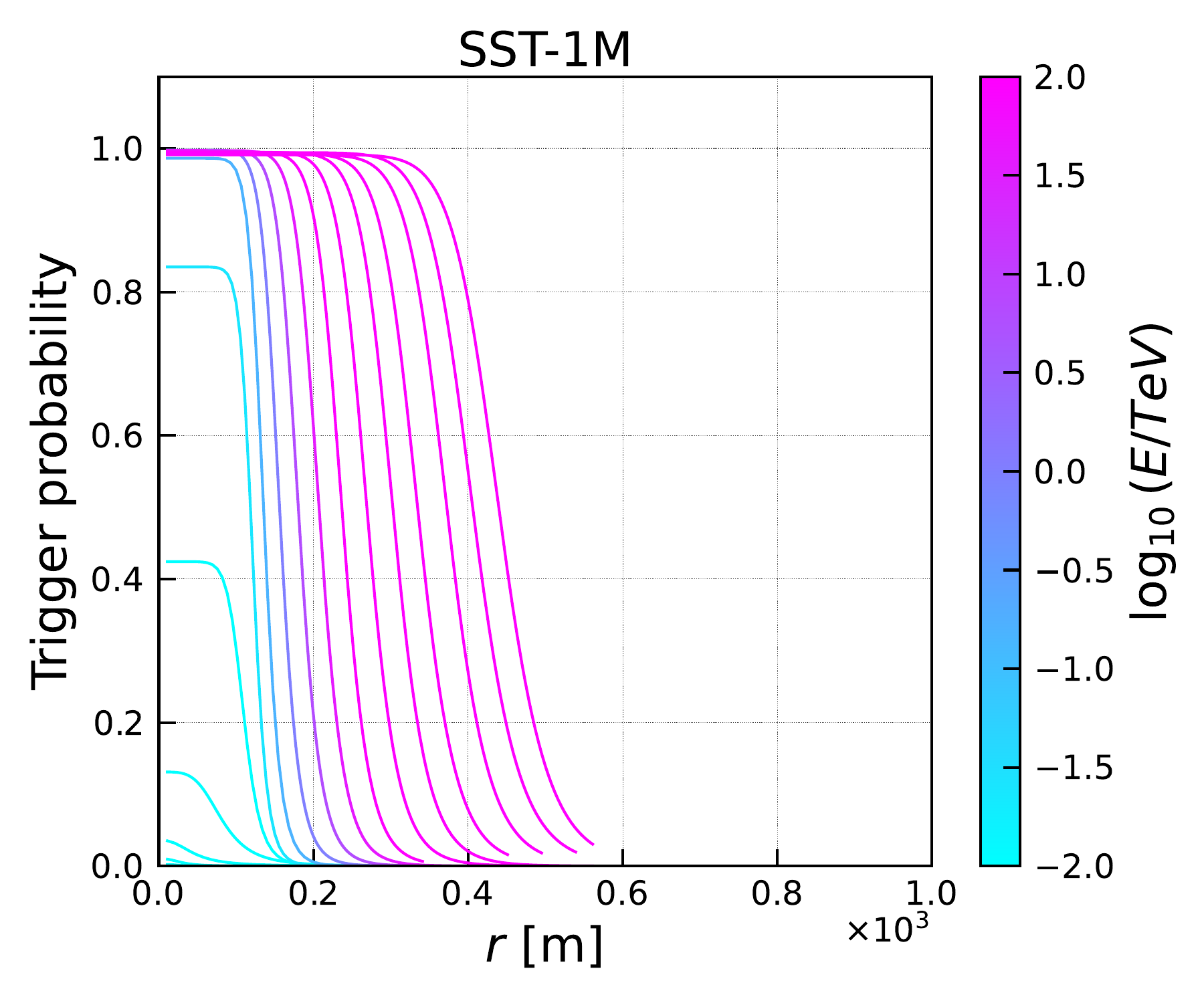}
    \includegraphics[width=0.3\linewidth]{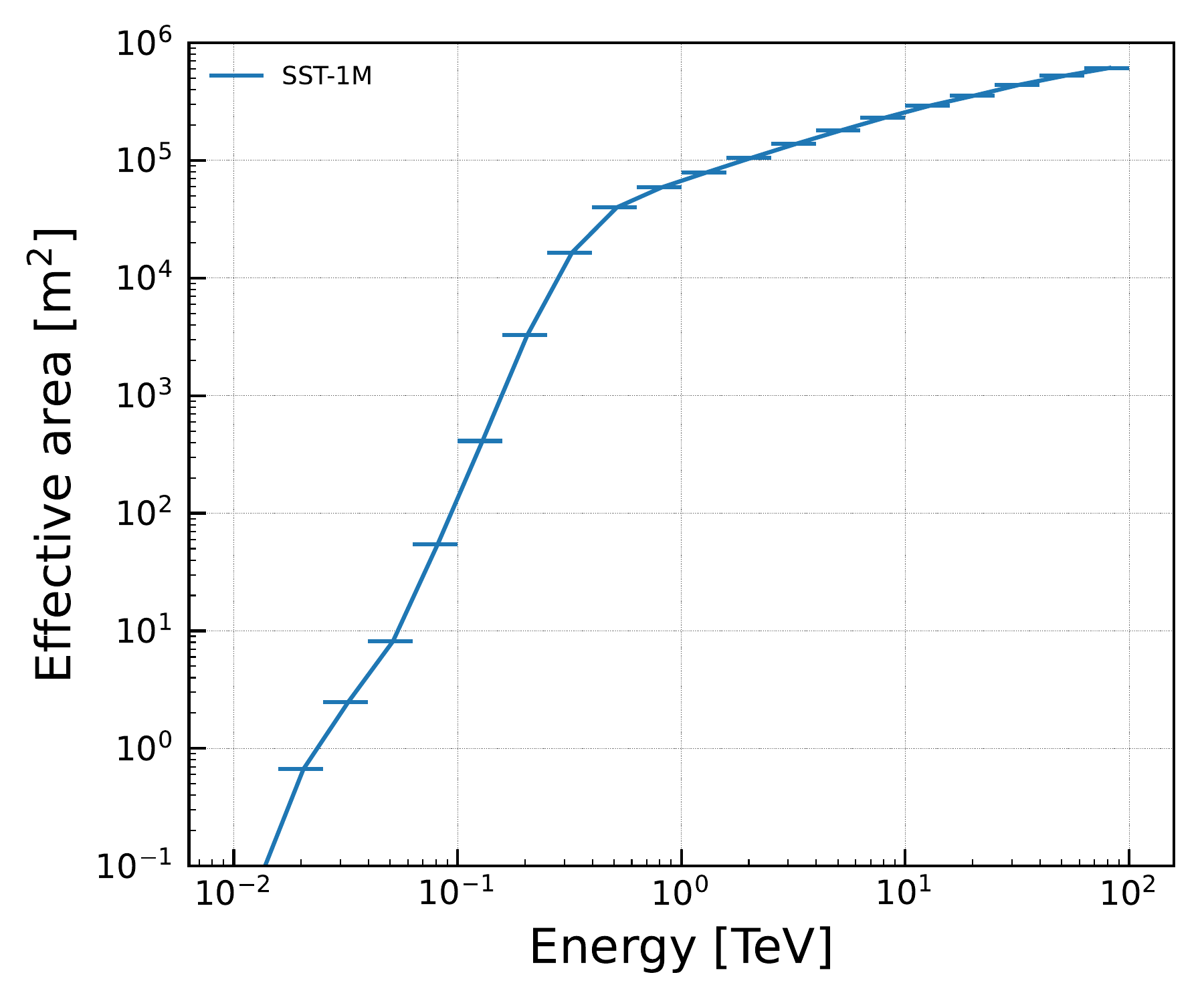}
    \caption{Left: Survival probability vs. impact distance for all energy bins. Center: Parametrized survival probability curves (Eq.~\ref{eq:mono_trigger_probability}). Right: Effective area computed from Eq.~\ref{eq:mono_effective_area}.}
    \label{fig:trigger_probability}
\end{figure}

The survival probability as a function of impact distance to the SST-1M telescope is shown in Figure~\ref{fig:trigger_probability}-left across all energy bins.

The survival probability $p(r, E)$ is modeled as:

\begin{equation}\label{eq:mono_trigger_probability}
    p(r, E) = \frac{p_{\rm max}(E)}{\left( \frac{r}{r_{1/2}(E)} \right)^{k(E)} + 1},
\end{equation}

where $p_{\rm max}(E) \leq 1$ is the maximum survival probability, $r_{1/2}(E)$ is the radius at which $p = 0.5$, and $k(E)$ controls the slope of the drop-off.

\begin{figure}
    \centering
    \includegraphics[width=0.3\linewidth]{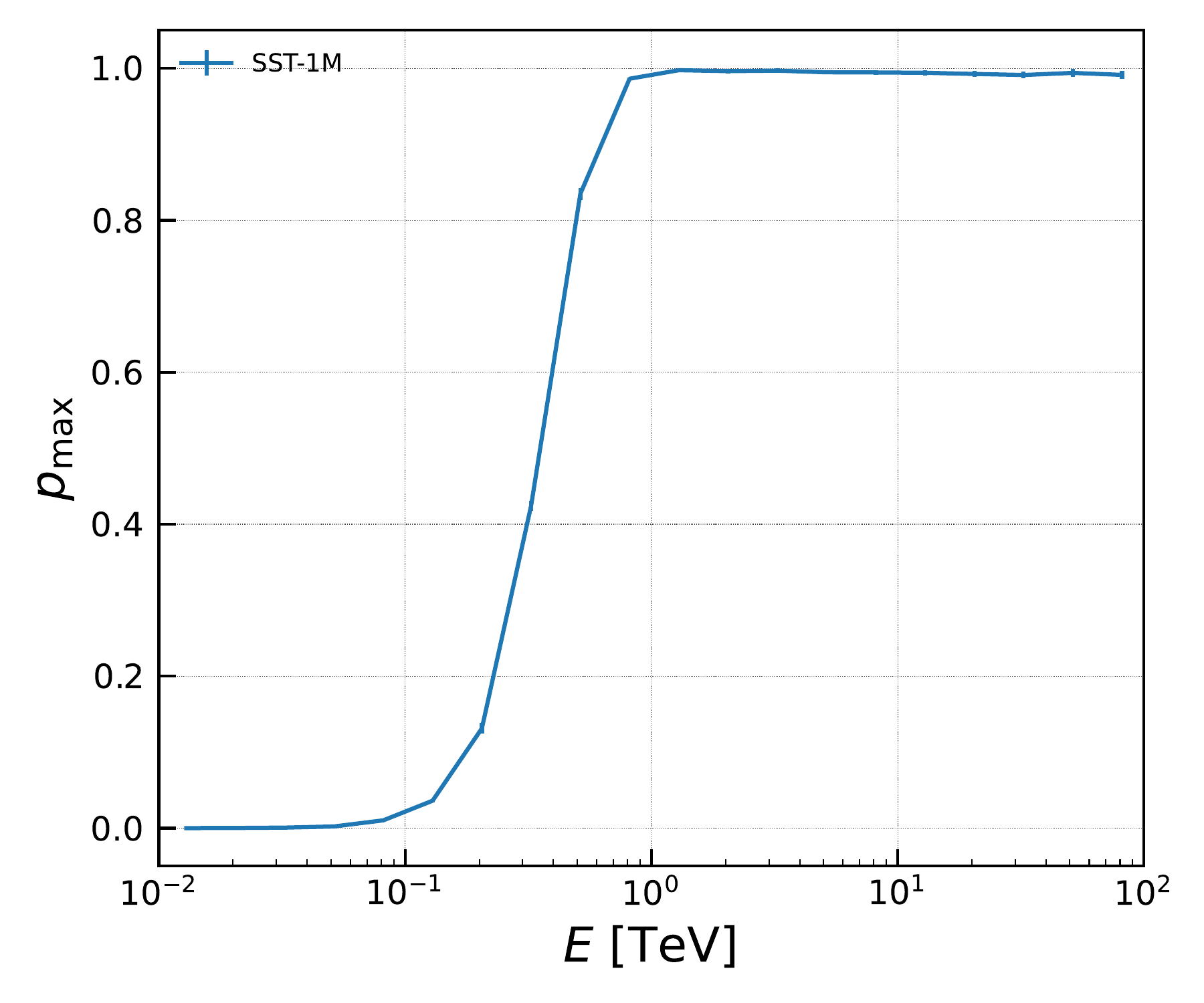}
    \includegraphics[width=0.3\linewidth]{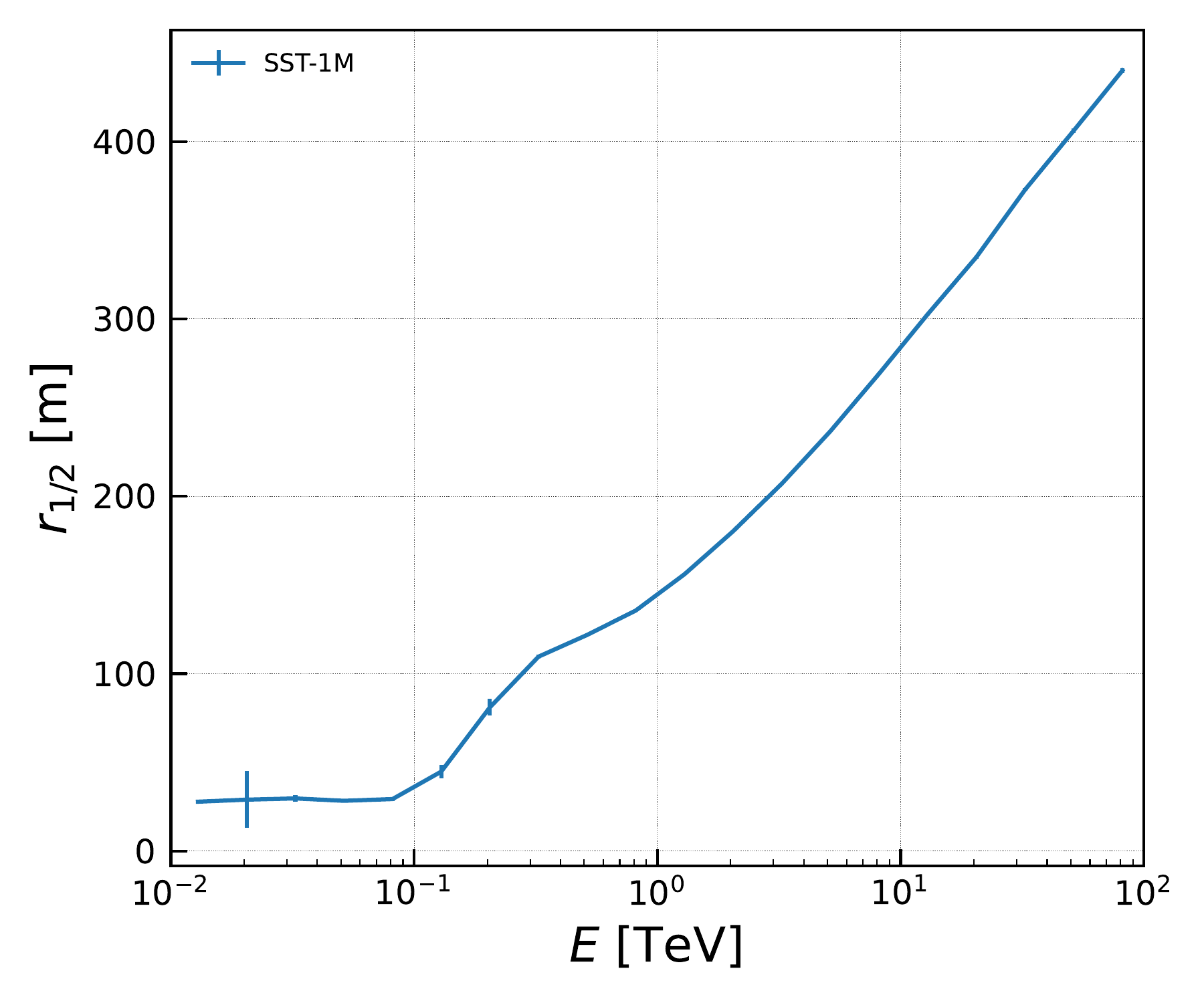}
    \includegraphics[width=0.3\linewidth]{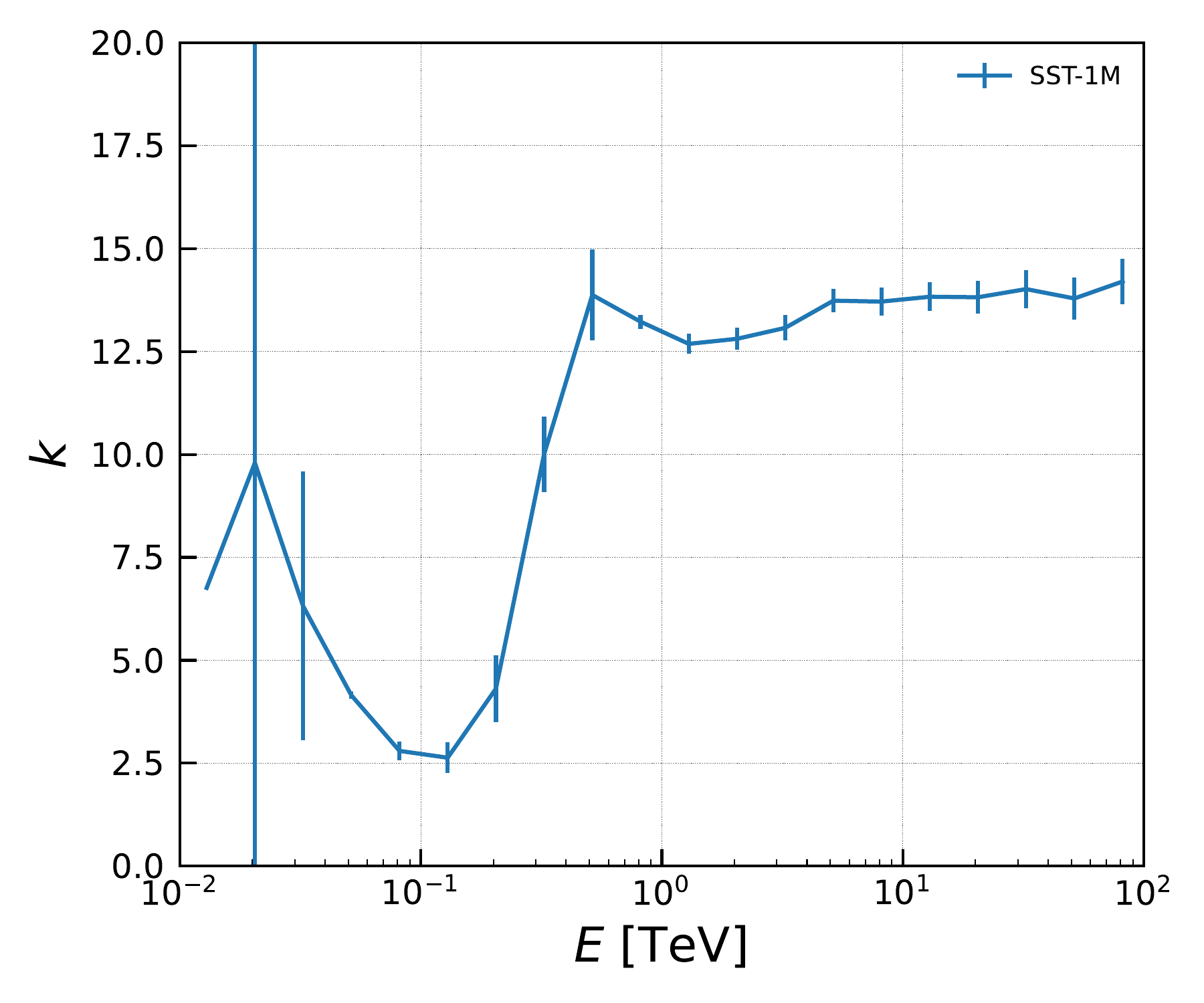}
    \caption{Survival probability parameters vs. energy: $p_{\rm max}(E)$ (left), $r_{1/2}(E)$ (center), and $k(E)$ (right).}
    \label{fig:fit_parameters}
\end{figure}

Fitted parameters as functions of energy are shown in Figure~\ref{fig:fit_parameters}.

The monoscopic effective area is computed by integrating the survival probability over the ground surface:

\begin{equation}\label{eq:mono_effective_area}
    A_{\rm eff}(E) = \int_0^{\infty} \int_0^{2\pi} p(r, E) \, r \, {\rm d}r \, {\rm d}\phi
\end{equation}

\subsubsection{Stereoscopic Effective Area}

For arrays with $N > 1$ telescopes, the stereoscopic effective area is derived from the joint survival probability. The probability that an event survives the trigger and the analysis cut with exactly $k$ telescopes is:

\begin{equation}\label{eq:stereo_prob}
    P(n = k) = \sum_{T_k \in T} \prod_{i \in T_k} p_i \prod_{i \notin T_k} (1 - p_i),
\end{equation}

where $T$ is the set of all telescopes and $T_k$ the set of $\binom{N}{k}$ combinations of $k$ telescopes among $N$ telescopes.

To define a valid array event, a minimum multiplicity $m$ is required. The probability that at least $m$ telescopes survive the trigger and analysis cuts is:

\begin{equation}\label{eq:stereo_multiplicity_prob}
    P(n \geq m) = 1 - \sum_{k=0}^{m-1} P(n = k).
\end{equation}

The stereoscopic effective area for multiplicity $m$ is then:

\begin{equation}\label{eq:stereo_effective_area}
    A_{\rm eff}^m(E) = \int_{-\infty}^{\infty} \int_{-\infty}^{\infty} P(n \geq m, x, y, E) \, {\rm d}x \, {\rm d}y.
\end{equation}

\begin{figure}
    \centering
    \includegraphics[width=0.3\linewidth]{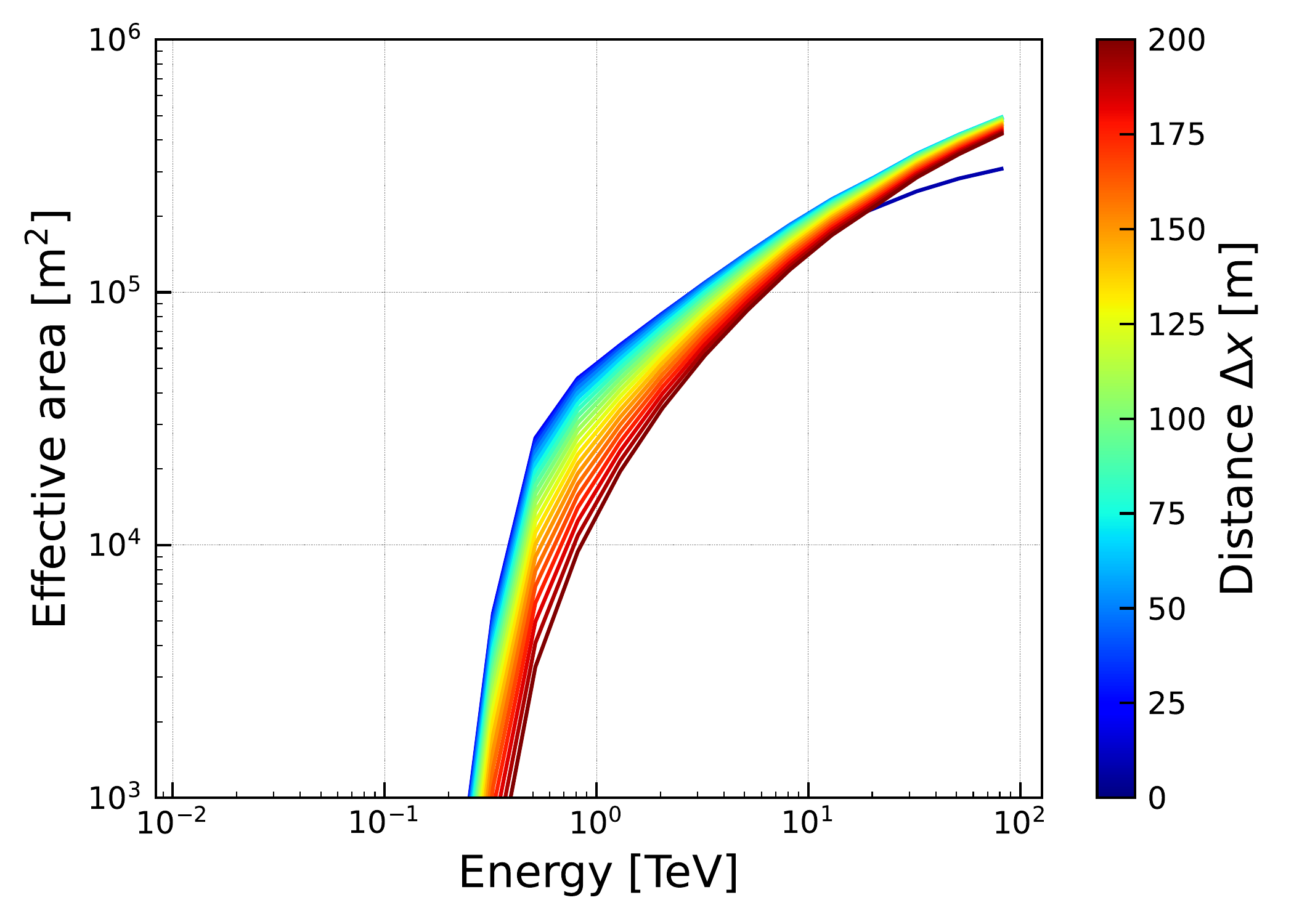}
    \includegraphics[width=0.3\linewidth]{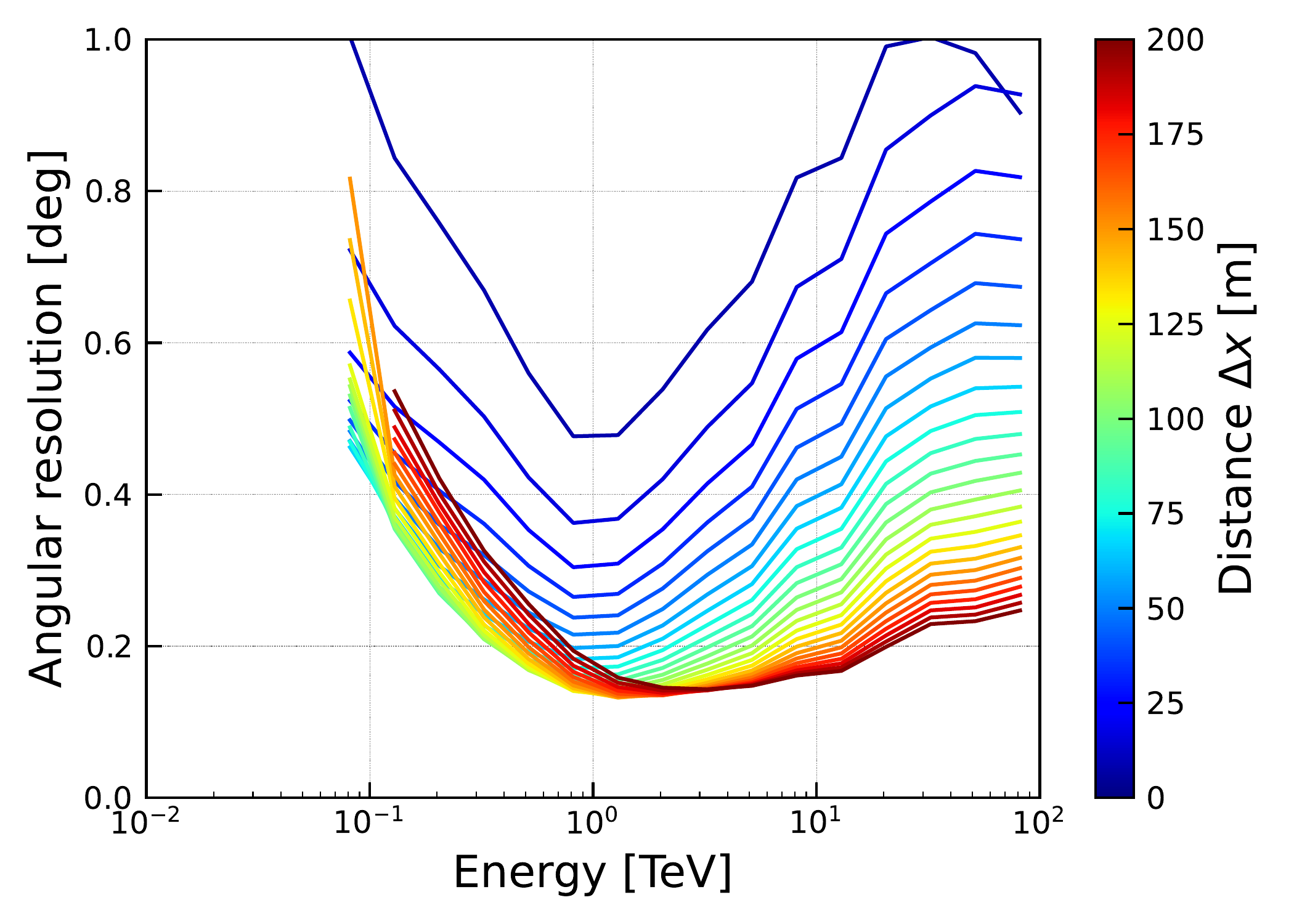}
    \includegraphics[width=0.3\linewidth]{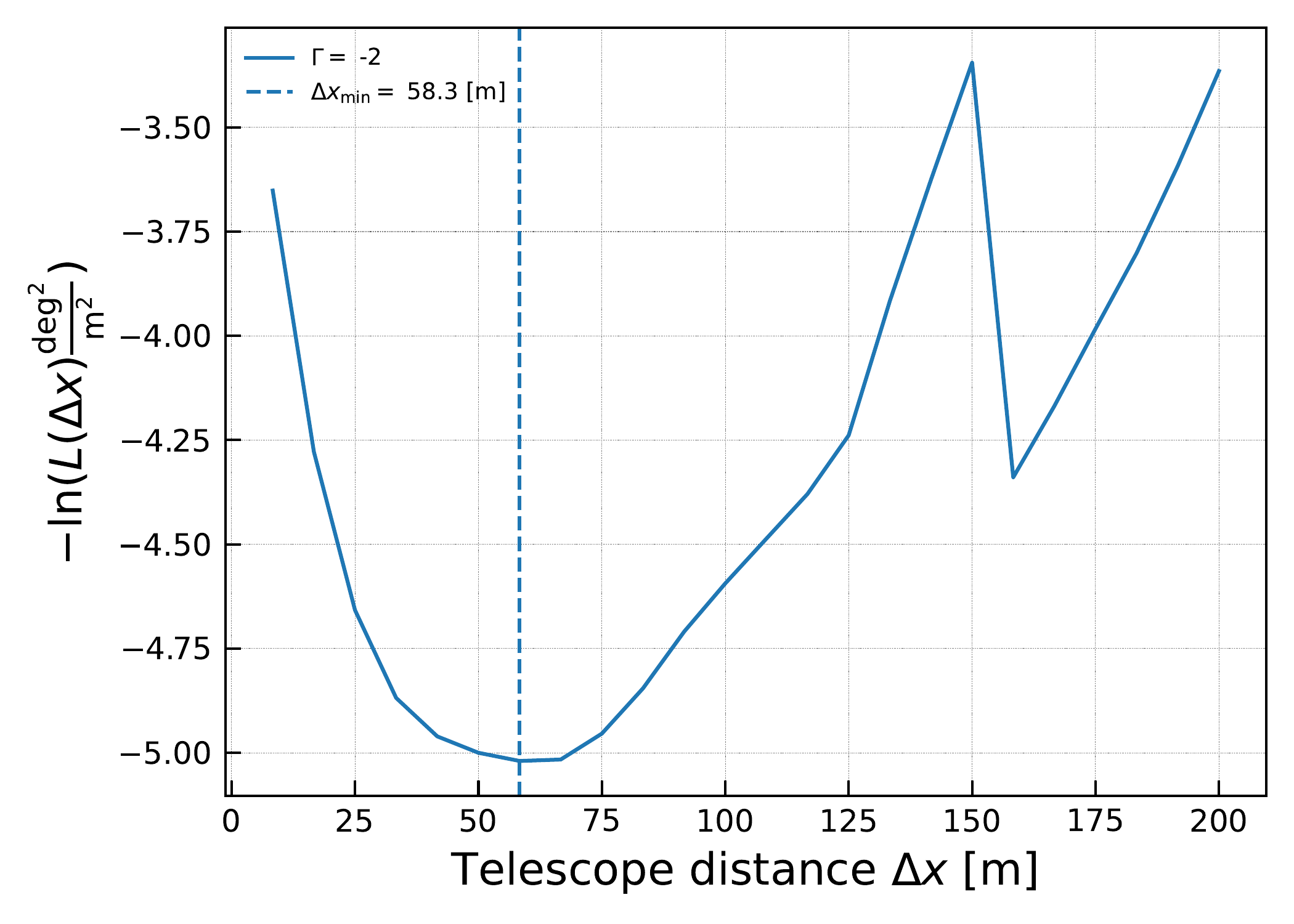}`
    \caption{Left: Effective area of a two-SST-1M array ($m=2$) vs. energy for various separations. Center: Angular resolution vs. energy and separation. Right: Cost function vs. separation $\Delta x$. A quality selection criteria of the stereo angle $\Delta \psi \geq 1^\circ$ was used in all figures.}
    \label{fig:stero_perf}
\end{figure}

Figure~\ref{fig:stero_perf}-left shows the effective area for an array with two-SST-1M telescopes (and multiplicity $m = 2$) across energy bins and various telescope separations. As expected, the effective area decreases with increasing separation.

\subsection{Angular Resolution}\label{sec:angular_resolution}

Direction reconstruction in IACTs can be performed using either monoscopic or stereoscopic techniques. While some algorithms allow reconstruction from a single telescope, others require multiple telescopes observing the same air shower. A comparison of stereoscopic methods are discussed in~\cite{Hofmann}.

We adopt the simplest reconstruction method: estimating the gamma-ray direction by intersecting the major axes of the images recorded by each telescope corresponding to algorithm 1 in~\cite{Hofmann}. Although more advanced techniques could improve precision, they are reserved for future work.

\subsubsection{Monoscopic Angular Resolution}

\begin{figure}
    \centering
    \includegraphics[width=0.25\linewidth]{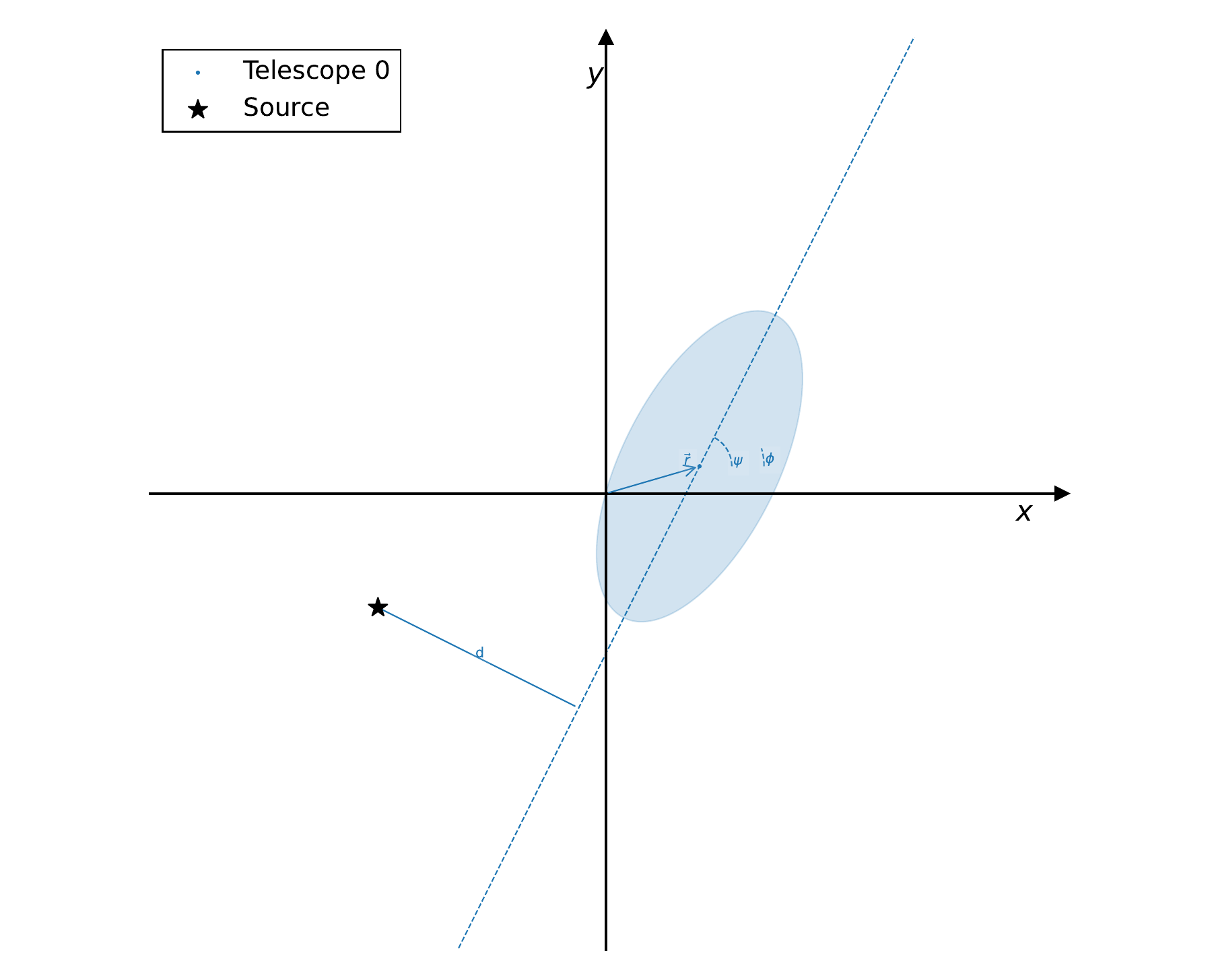}
    \includegraphics[width=0.3\linewidth]{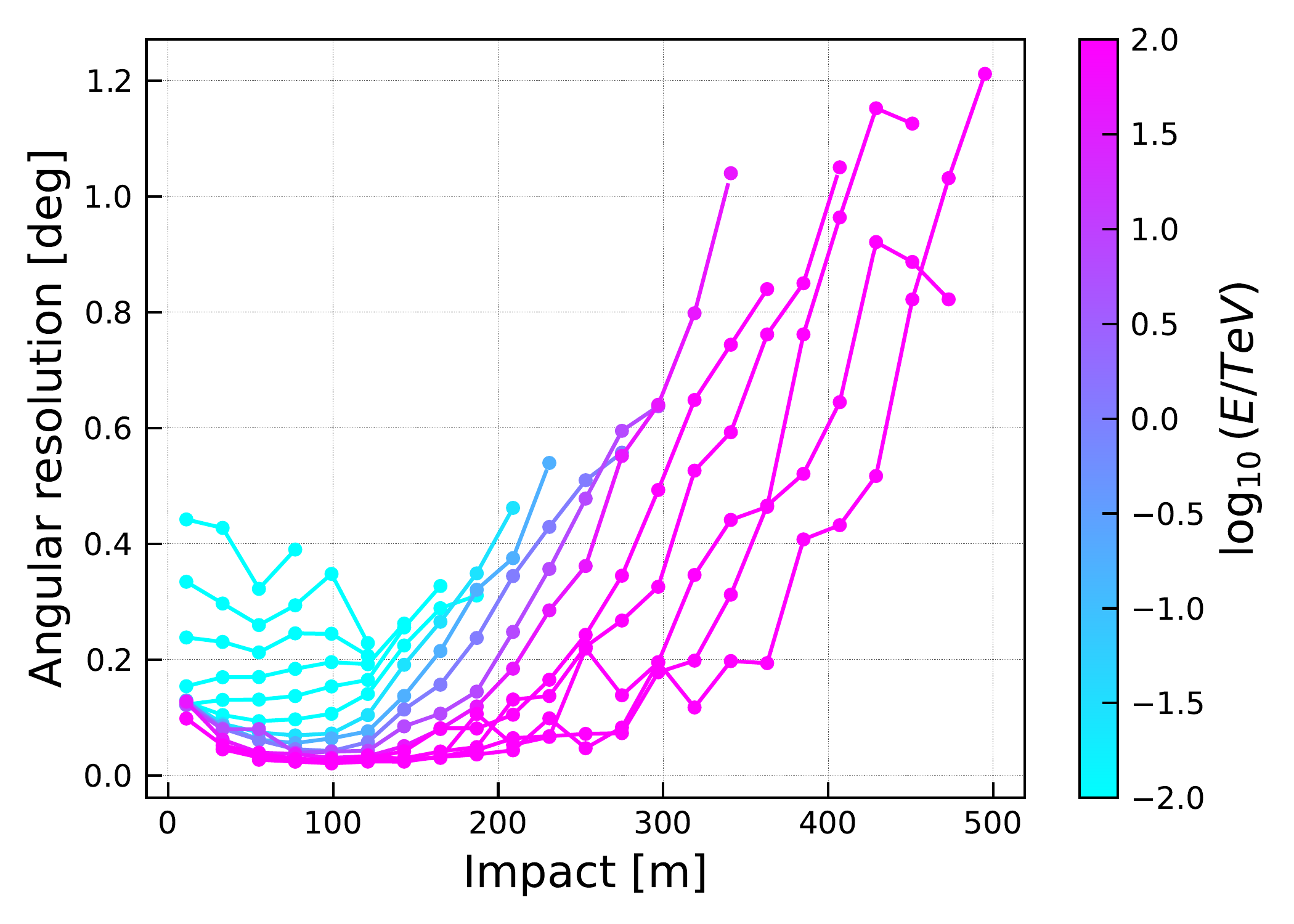}
    \caption{Left: Illustration of the "miss" distance $d$ between the source direction and the major axis of the image. Right: Standard deviation $\sigma_{\theta}$ of the monoscopic angular separation as a function of impact distance for different energy bins.}
    \label{fig:mono_angular_resolution}
\end{figure}

With the selected reconstruction method, the direction of the primary particle cannot be fully determined from a single telescope. However, the shortest distance from the true source position to the major axis, known as the "miss" parameter, can be computed (see Figure~\ref{fig:mono_angular_resolution}-left):

\begin{equation}\label{eq:mono_distance}
    d = \cos \psi \, (y_{\rm{cm}} - y^{*}) - \sin \psi \, (x_{\rm{cm}} - x^{*}),
\end{equation}

where $\psi$ is the angle of the major axis, $(x_{\rm cm}, y_{\rm cm})$ is the image centroid, and $(x^{*}, y^{*})$ is the true direction of the primary gamma ray. In the literature, this parameter $d$ is sometimes referred to as the "miss" parameter, representing how much the reconstructed event deviates from the true source position.

This distance $d$ is converted to angular units using the telescope focal length $F$:

\begin{equation}\label{eq:mono_angular_distance}
    \theta = \arctan \left( \frac{d}{F} \right) \simeq \frac{d}{F}.
\end{equation}

The angular separation $\theta$ between the source direction and the major axis was computed for all events surviving the analysis cut and trigger. Its standard deviation $\sigma_{\theta}$ as a function of the impact distance to the telescope is shown in Figure~\ref{fig:mono_angular_resolution}-right for all energy bins.

The monoscopic angular resolution $\sigma_{\theta}$ for the SST-1M telescope, shown in Figure~\ref{fig:mono_angular_resolution}-right, has not been parametrized as was done previously in Section~\ref{sec:aeff}. This parametrization will be addressed in future work. For now, piecewise linear interpolation is used.

\subsubsection{Stereoscopic Angular Resolution}

With $k > 1$ telescopes, the reconstructed direction $\vec{x}$ minimizing the sum squared distance, $\sum_{i=0}^{k-1} d_i^2$, to the $k$ major axis lines is:

\begin{equation}\label{eq:stereo_resolution}
    \underset{\vec{x} \in \mathbb{R}^2}{\rm arg\,min} \sum_{i=0}^{k-1} (\vec{x} - \vec{x}_{{\rm cm},i})^\intercal (\hat{n}_i \hat{n}_i^\intercal) (\vec{x} - \vec{x}_{{\rm cm},i}),
\end{equation}

where $\hat{n}_i = (-\sin \psi_i, \cos \psi_i)$ is the unit vector perpendicular to the major axis of telescope $i$, oriented at angle $\psi_i$, and $\vec{x}_{{\rm cm},i} = (x_{{\rm cm},i}, y_{{\rm cm},i})$ is the centroid of the image in telescope $i$. 

The analytical solution to Equation~\ref{eq:stereo_resolution} is:

\begin{equation}
    \vec{x} = \left( \sum_{i=0}^{k-1} \hat{n}_i \hat{n}_i^\intercal \right)^{-1} \left( \sum_{i=0}^{k-1} \hat{n}_i \hat{n}_i^\intercal \vec{x}_{{\rm cm},i} \right).
\end{equation}

Assuming that each telescope yields an error $\sigma_{\theta_i}$ in the direction $\hat{n}_i$, the angular error on the reconstructed direction $\sigma_{\theta}$ (from the generalized variance) can be expressed as:

\begin{equation}\label{eq:angular_resolution_stereo}
    \frac{1}{\sigma_{\theta}^2} =  \sum_{i<j}^{k-1}  \frac{1}{\sigma_{\theta_i} \sigma_{\theta_j}} \left| \sin \Delta \psi_{ij} \right|, \quad \Delta \psi_{ij} = \psi_i - \psi_j.
\end{equation}

\begin{figure}
    \centering
    \includegraphics[width=0.3\linewidth]{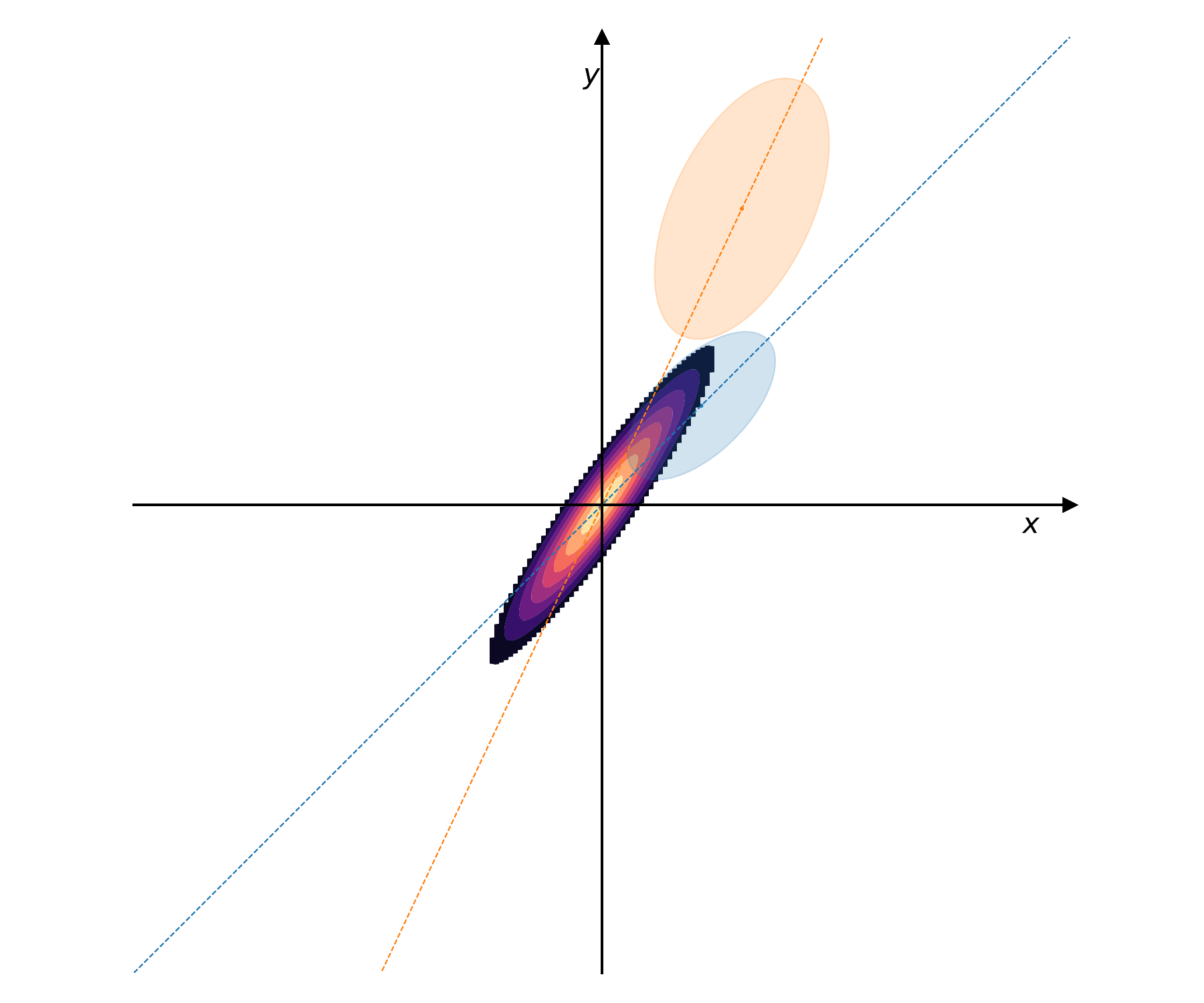}
    \includegraphics[width=0.3\linewidth]{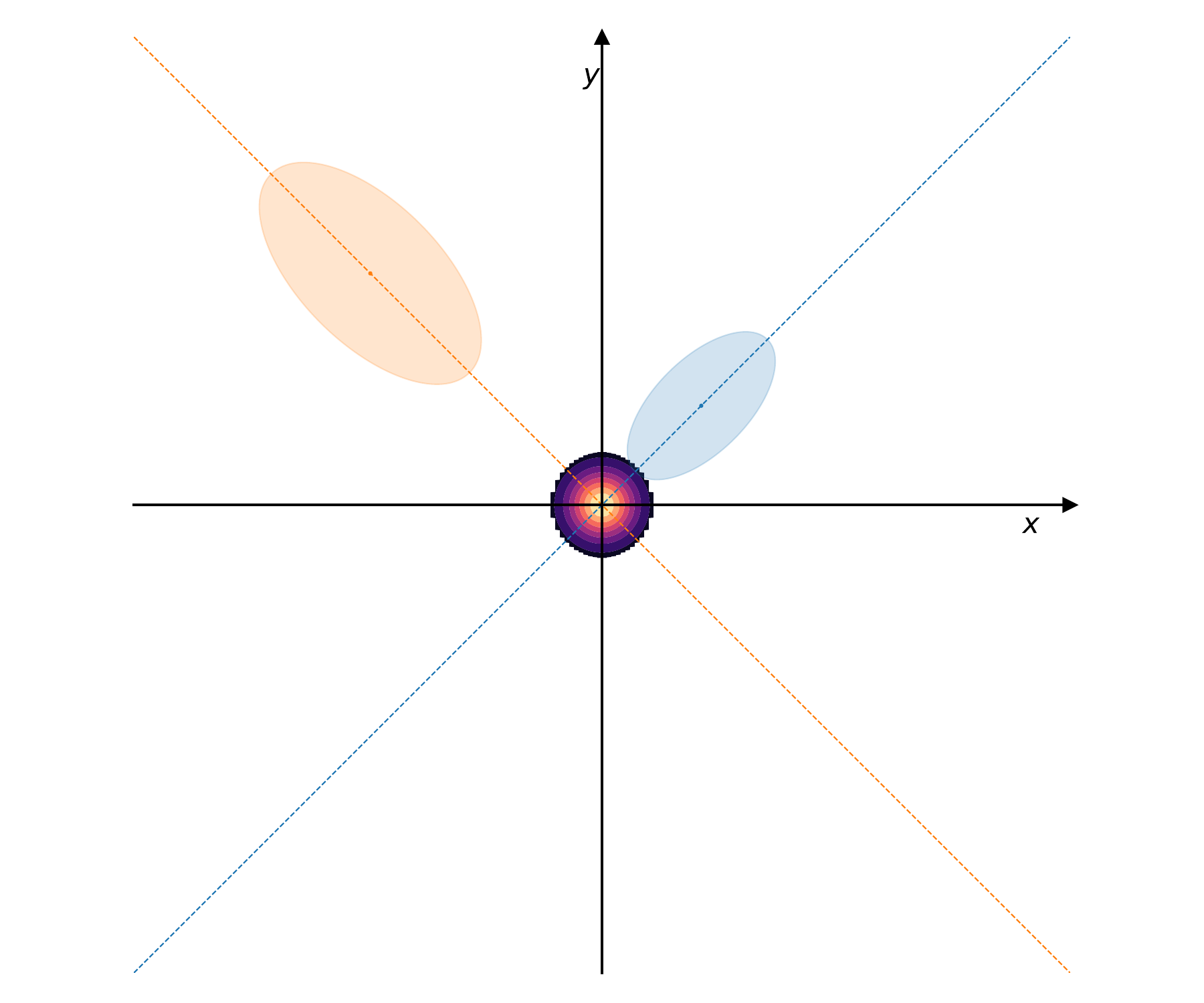}
    \includegraphics[width=0.3\linewidth]{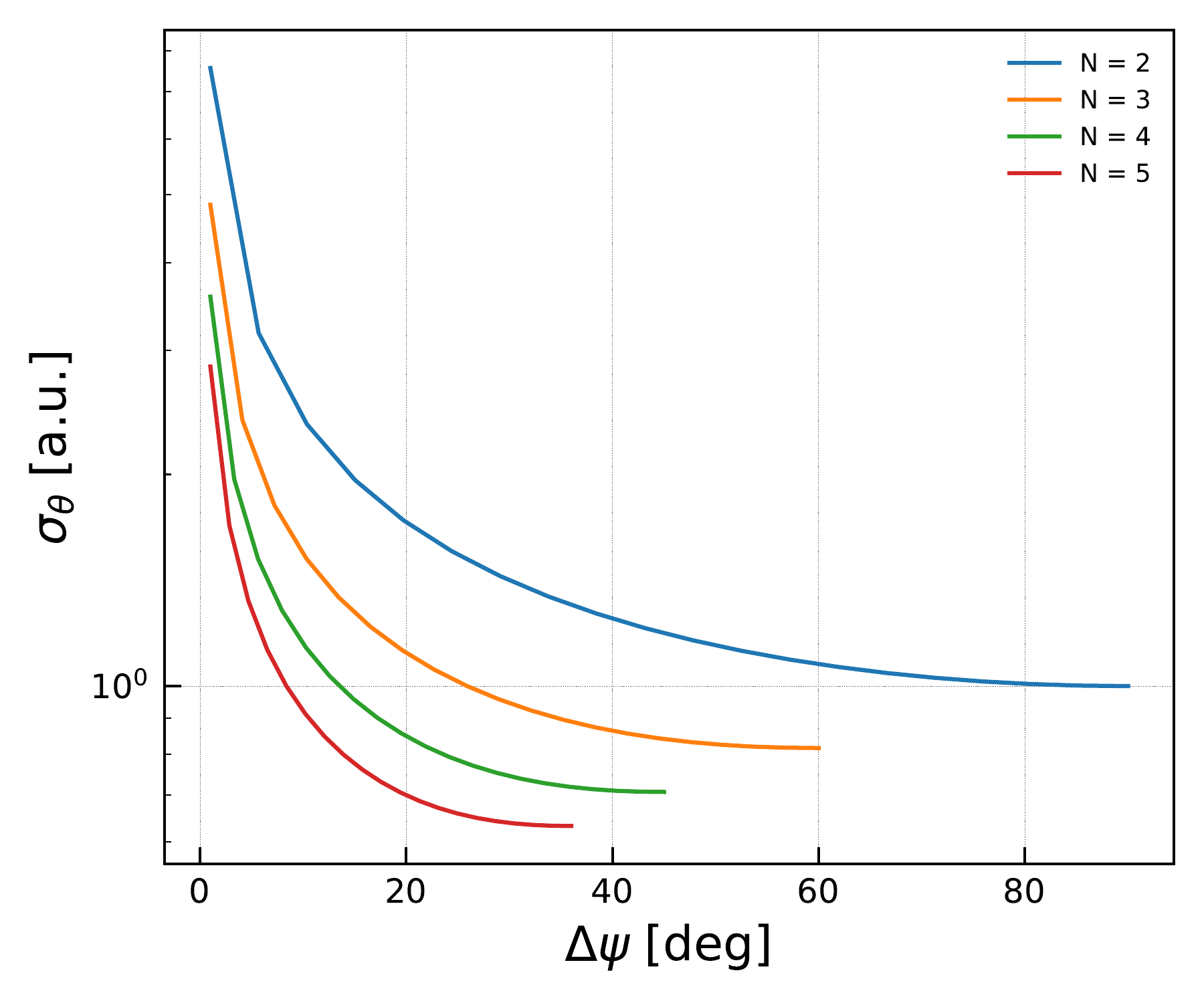}
    \caption{Left: Poorly constrained reconstruction due to parallel major axes. Center: Optimal reconstruction with perpendicular major axes. Right: Improvement of angular resolution with increasing separation angle $\Delta \psi$ and number of telescopes.}
    \label{fig:stereo_angular_scheme}
\end{figure}

From Equation~\ref{eq:angular_resolution_stereo}, we observe that the stereoscopic angular resolution:

\begin{itemize}
    \item Is proportional to the monoscopic angular resolutions,
    \item Improves with the number of telescope pairs combination ($\propto \sqrt{\frac{2}{k(k-1)}}$), see Figure~\ref{fig:stereo_angular_scheme}-right,
    \item Depends on the angular separation $\Delta \psi_{ij}$, sometimes called "stereo angle", between the major axes of telescope pairs.
\end{itemize}

This last point reflects a geometric constraint: if the major axes tend to be parallel ($|\sin \Delta \psi |\simeq 0$), as illustrated in Figure~\ref{fig:stereo_angular_scheme}-left, the reconstructed direction is poorly constrained. Conversely, if they are nearly perpendicular ($|\sin \Delta \psi| \simeq 1$), as shown in Figure~\ref{fig:stereo_angular_scheme}-center, the reconstruction is optimal. Moreover, the reconstruction is impossible at $\sin \Delta \psi = 0$ and thus it is necessary to apply quality cut on the stereo angle. In this work, we selected event with $\Delta \psi \geq 1^\circ$.

An additional important remark is that the average angle $\psi$ of the major axes aligns with the polar angle $\phi$ of the impact point. Therefore, Equation~\ref{eq:angular_resolution_stereo} provides a way to compute the stereoscopic angular resolution from the monoscopic angular resolutions of individual telescopes, for any impact point of the primary particle.

The average angular resolution of a pair of SST-1M telescopes for various separation distances is shown in Figure~\ref{fig:stero_perf}-center across all energy bins. The average was computed over the full surface, weighted by the array event survival probability.

\section{Layout Optimization}\label{sec:optimization}

Having constructed a differentiable model for the stereoscopic performance metrics—namely the effective area and angular resolution—based on monoscopic performance derived from Monte Carlo simulations, we can now proceed to optimize the array layout. As previously discussed, we focus on a simplified configuration consisting of only two SST-1M telescopes. This choice offers the advantage of reducing the optimization problem to a single free parameter: the telescope separation.

The stereoscopic effective area and angular resolution are shown in Figure~\ref{fig:stero_perf} (left and center) across all energy bins, for a range of tested separations from 0 to 200 meters. 

For the optimization, we consider a science case aimed at detecting a gamma-ray source with the highest possible statistical significance, assuming a spectral index of $\Gamma = -2$, while also achieving the sharpest possible image resolution. To this end, we define a cost function:

\begin{equation}\label{eq:cost} L(\Delta x) = - \frac{\sum_k A_{\rm{eff}}^{m}(E_k, \Delta x) w_k}{\sum_k \sigma^2_{\theta}(E_k, \Delta x) w_k}, \quad w_k \sim E^{\Gamma}, \quad \Gamma = -2, \end{equation}

where $A_{\rm{eff}}^{m}$ is the effective area for multiplicity $m=2$, $\sigma_{\theta}$ is the angular resolution, and $w_k$ are energy-dependent weights reflecting the assumed source spectrum. The angular resolutions and effective areas are computed accordingly with the selected quality cut $\Delta \psi \geq 1^\circ$.

The resulting loss function $L$ as a function of telescope separation is shown in Figure~\ref{fig:stero_perf}-right. The optimal separation, corresponding to the minimum of $L$, is found to be at $\Delta x_{{\rm min}} = 58.3$~m.

\section{Conclusion and Outlooks}\label{sec:conclusion}

In this study, we developed a differentiable model to estimate the stereoscopic performance metrics (effective area and angular resolution) of an Imaging Atmospheric Cherenkov Telescope (IACT) array, based on monoscopic performance derived from Monte Carlo simulations. Focusing on a simplified configuration of two identical SST-1M telescopes located at the Hanle Observatory, we demonstrated how these metrics vary with telescope separation and energy.

To identify an optimal array layout, we introduced a cost function that balances sensitivity (via effective area) and image sharpness (via angular resolution), tailored to a science case targeting a gamma-ray source with a spectral index of $\Gamma = -2$. By evaluating this cost function across a range of telescope separations, we found that the optimal configuration occurs at a separation of 66.7~m.

This framework provides a foundation for scalable optimization of more complex arrays and can be extended to include additional performance metrics such as energy resolution and background rejection. Future work will focus on refining the angular resolution model, including signal extraction from the dominant hadronic background, adding energy reconstruction, incorporating more advanced reconstruction algorithms, and generalizing the optimization to large heterogeneous telescope arrays. Moreover using differential programming would greatly improve the computational effectiveness of finding the optimized the array layout as suggested in~\cite{MODE}. Finally in the future we will optimize the layout based on the quality cuts. Here we selected a single fixed quality threshold on the stereo angle but quality cut criterion (such as on image intensity and on stereo angle) could be simultaneously optimized with the layout.

\bibliographystyle{JHEP} 
\bibliography{bibliography}

\acknowledgments

\setlength{\baselineskip}{10pt}
{\scriptsize This work was conducted in the context of investigating the implementation of an IACT array at the Hanle Observatory. The authors thank our colleagues from the MACE and the SST-1M collaborations for their fruitful input and support. Special thanks are addressed to Mohan Karthik for producing the Monte Carlo simulations of the SST-1M telescope at the Hanle cite. This publication was created as part of the projects funded in Poland by the Minister of Science based on agreements number 2024/WK/03 and DIR/\-WK/2017/12. The construction, calibration, software control and support for operation of the SST-1M cameras is supported by SNF (grants CRSII2\_141877, 20FL21\_154221, CRSII2\_160830, \_166913, 200021-231799), by the Boninchi Foundation and by the Université de Genève, Faculté de Sciences, Département de Physique Nucléaire et Corpusculaire. The Czech partner institutions acknowledge support of the infrastructure and research projects by Ministry of Education, Youth and Sports of the Czech Republic (MEYS) and the European Union funds (EU), MEYS LM2023047, EU/MEYS CZ.02.01.01/00/22\_008/0004632, CZ.02.01.01/00/22\_010/0008598, Co-funded by the European Union (Physics for Future – Grant Agreement No. 101081515), and Czech Science Foundation, GACR 23-05827S.}

\end{document}